\documentclass{article}

\usepackage{amsmath}    %
\usepackage{graphicx}   %

\usepackage{amssymb,amsfonts,mathrsfs}
\usepackage{mathtools}
\usepackage[amsmath,thmmarks]{ntheorem}
\usepackage{dsfont}
\usepackage{colortbl}
\usepackage{algorithmic}
\usepackage{multirow,booktabs,bigdelim}
\usepackage{caption}
\usepackage{algorithm}
\usepackage{comment}
\usepackage{natbib}
\usepackage{siunitx}
\usepackage{url}
\usepackage{paralist}
\usepackage{enumitem}
\usepackage{tikz}
\usepackage{tkz-graph}
\usetikzlibrary{shapes.geometric}
\usetikzlibrary{backgrounds}
\usetikzlibrary{arrows}
\usepackage[framemethod=TikZ]{mdframed}
\usepackage{dirtytalk}
\usepackage[labelsep=period]{caption}
\usepackage[hidelinks]{hyperref}
\usepackage{graphicx}
\newdimen\arrowsize
\pgfarrowsdeclare{arcsq}{arcsq}
{
  \arrowsize=0.2pt
  \advance\arrowsize by .5\pgflinewidth
  \pgfarrowsleftextend{-4\arrowsize-.5\pgflinewidth}
  \pgfarrowsrightextend{.5\pgflinewidth}
}
{
  \arrowsize=1.5pt
  \advance\arrowsize by .5\pgflinewidth
  \pgfsetdash{}{0pt} %
  \pgfsetroundjoin   %
  \pgfsetroundcap    %
  \pgfpathmoveto{\pgfpoint{0\arrowsize}{0\arrowsize}}
  \pgfpatharc{-90}{-140}{4\arrowsize}
  \pgfusepathqstroke
  \pgfpathmoveto{\pgfpointorigin}
  \pgfpatharc{90}{140}{4\arrowsize}
  \pgfusepathqstroke
}

\newcommand{\B}[1]{\mathbf{#1}}

\newcommand{\eps}{\varepsilon}  
\newcommand{\tPA}[2][]{{\widetilde{\B{PA}}}^{#1}_{#2}}\newcommand{\PA}[2][]{{\B{PA}}^{#1}_{#2}}

\newcommand{\doo}[1]{{do\left({#1}\right)}}
\newcommand{\dott}[1]{{\tfrac{\text{d}}{\text{d}t}{#1}}}
\newcommand{\dooo}[1]{{do({#1})}}

\newboolean{ARXIVVER}
\setboolean{ARXIVVER}{true}

\ifthenelse {\boolean{ARXIVVER}}
{
\usepackage{fullpage}
}
{
}

\title{Causal models for dynamical systems}
\author{Jonas Peters \\
University of Copenhagen, Denmark \\
\url{jonas.peters@math.ku.dk}
\and Stefan Bauer \\
MPI T\"ubingen, Germany\\
\url{stefan.bauer@tuebingen.mpg.de}
\and Niklas Pfister\\
University of Copenhagen, Denmark\\
\url{np@math.ku.dk}
}
\date{\today}

\begin{document}

\ifthenelse {\boolean{ARXIVVER}}
{
\maketitle

\abstract{
A probabilistic model describes a system in its observational state. 
In many situations, however, we are interested in the system's response under interventions. 
The class of structural causal models provides a language that allows us to model the behaviour under interventions. 
It can been taken as a starting point to answer a plethora of causal questions, including the identification of causal effects or causal structure learning.
In this chapter, we provide a natural and straight-forward extension 
of this concept to dynamical systems, 
focusing on continuous time models. 
In particular, we introduce 
two types of causal kinetic models that differ in how the randomness enters into the model: it may either be considered as observational noise or as systematic driving noise. 
In both cases, we define interventions and therefore provide a possible starting 
point for causal inference. In this sense, the book chapter 
provides more questions than answers.
The focus of the proposed causal kinetic models lies on
the dynamics themselves rather than corresponding stationary distributions, for example. 
We believe that this is beneficial 
when the aim is to model the full time evolution of the
system
and data are measured at different time points. 
Under this focus, it is natural to consider
interventions in the differential equations themselves. 
}
}
{
\chapter{%
Causal models for dynamical systems
}
\label{sec:causal_models}

\begin{center}
Jonas Peters (University of Copenhagen)\\
Stefan Bauer (MPI T\"ubingen)\\
Niklas Pfister (ETH Zurich)
\end{center}

\begin{mdframed}[roundcorner=5pt, frametitle={Abstract}]
A probabilistic model describes a system in its observational state. 
In many situations, however, we are interested in the system's response under interventions. 
The class of structural causal models provides a language that allows us to model the behaviour under interventions. 
It can been taken as a starting point to answer a plethora of causal questions, including the identification of causal effects or causal structure learning.
In this chapter, we provide a natural and straight-forward extension 
of this concept to dynamical systems, 
focusing on continuous time models. 
In particular, we introduce 
two types of causal kinetic models that differ in how the randomness enters into the model: it may either be considered as observational noise or as systematic driving noise. 
In both cases, we define interventions and therefore provide a possible starting 
point for causal inference. In this sense, the book chapter 
provides more questions than answers.
The focus of the proposed causal kinetic models lies on
the dynamics themselves rather than corresponding stationary distributions, for example. 
We believe that this is beneficial 
when the aim is to model the full time evolution of the
system
and data are measured at different time points. 
Under this focus, it is natural to consider
interventions in the differential equations themselves. 
\end{mdframed}
}

We wholeheartedly congratulate Judea Pearl on winning the Turing Award.
His groundbreaking work has inspired much of our work, with this book chapter being
only one of several examples.

\section{Introduction}
In causality, we aim to understand how a system reacts under interventions, e.g., 
in gene knock-out experiments. 
There are different interventions we
can perform (including none at all), and we therefore 
require different 
descriptions of the data generating process. 
Some systems may be adequately described by 
deterministic equations, but  
if the system possesses
observational noise, 
unobserved factors or intrinsic randomness,
data generating processes are more appropriately modeled using the language of probability.
In data-driven sciences, we are used to model the data generating process 
with a single probability distribution, e.g., 
using a multivariate Gaussian 
with a certain covariance matrix. 
As argued above, however, causal models come with a plethora of distributions:
one distribution for each type of 
modeled
 intervention.
In general, the intervention distributions
are not arbitrarily different as it would be meaningless to talk
about a single underlying system otherwise.
It is a key challenge to describe which parts
of the distribution change and which parts remain invariant when 
considering different interventions.
Many researchers from various disciplines engaged in 
this question and developed the fundamental assumptions 
that are often referred to as invariance, autonomy or
modularity \citep{Wright1921, Haavelmo1944, Aldrich1989, Hoover1990,
Imbens2015, richardson2013single}. 
The concept of invariance relies heavily on what it means to
intervene on a system, making a precise formulation of interventions
crucial for causal modeling. Arguably one of the clearest formulation
of interventions is Judea Pearl's $do$-formalism
\citep[Chapter~1]{Pearl2009}.
One starts with a fixed reference distribution called the
observational distribution; 
one may think of it as describing the system
in its natural state with no intervention being performed. 
The system and its corresponding distribution
is assumed to have a modular structure,
and performing a
$do$-intervention means 
changing some of the modules.
This process yields an 
intervention distribution, often denoted 
by a $do(.)$
subscript. 
While this description can be made formal 
in various ways, we focus on one
that is based on structural causal models \citep{Wright1921,
  Bollen1989, Pearl2009}.  Usually, the formulation of structural
causal models, or SCMs, include random variables.  We believe,
however, that the descriptive power of SCMs lies in their modular
structure, which can be separated from randomness.  We therefore
introduce two different versions of SCMs: a deterministic version with
measurement noise and a version containing random variables.

\subsection{Structural causal models with measurement noise}\label{sec:SCM_measurenoise}
A \emph{deterministic structural causal model} (SCM) over $d$ variables
$x_1, \ldots, x_d$ is a collection of $d$ assignments
\begin{align} \label{eq:scmiiddet}
x^k &:= f^k(x^{\PA[]{k}}), \quad k=1, \ldots, d,
\end{align}
where for any $k \in \{1, \ldots, d\}$, 
$\PA[]{k} \subseteq \{1, \ldots, d\} \setminus \{k\}$ is called the
set of direct parents of $x^k$, and
$f^k$ is a real-valued function.
If $\PA[]{k} = \emptyset$, then
$f^k(x^{\PA[]{k}})$ should be interpreted as 
a constant.
For each SCM, we obtain a
corresponding graphical representation of the causal structure over the vertices\footnote{By slight abuse of
  notation, we identify $(x^1, \ldots, x^d)$ with its indices
  $(1, \ldots, d)$.} 
  $(1, \ldots, d)$
  by drawing directed edges from
$\PA[]{k}$ to $k$ for all $k \in \{1, \ldots, d\}$.
We further assume that the
system~\eqref{eq:scmiiddet}
is uniquely solvable,
which may be the case, even if the graph contains directed cycles, such as 
$3 \rightarrow 1 \rightarrow 4 \rightarrow 3$.
Each SCM then induces a state of the system
characterized 
by a single point in $\mathbb{R}^d$.
We will see in Section~\ref{sec:interventionsiid}  %
that the modular structure of~\eqref{eq:scmiiddet} is key 
to 
the ability to serve as a causal model. 
The assignments in~\eqref{eq:scmiiddet} can be thought of as lines in a computer program
that generate a specific state of the system. 
Interventions will be modeled as replacements 
of some of these lines.

We may now assume to  obtain noisy observations of the system, e.g., 
for each $k \in \{1, \ldots, d\}$,
we may have
\begin{equation} \label{eq:obsnoiseiid}
X^k := x^k + \boldsymbol{\eps}^k,
\end{equation}
where 
$
\boldsymbol{\eps}^1,
\ldots,
\boldsymbol{\eps}^d
$
are jointly independent random variables. 
Instead of a single point, this model now induces a joint distribution over the 
observed random variables $X^1, \ldots, X^d$.

\subsection{Structural causal models with driving noise}\label{sec:SCM_drivingnoise}
More common than the above approach is the assumption that the
randomness enters inside the structural assignments. Formally, a
\emph{stochastic structural causal model} over $d$ random variables
$X_1, \ldots, X_d$ is a collection of $d$ assignments
\begin{align} \label{eq:scmiid}
X^k &:= f^k(X^{\PA[]{k}}, \varepsilon^k), \quad k=1, \ldots, d,
\end{align}
together with a distribution over the noise variables
$\varepsilon^1, \ldots, \varepsilon^d$. As above, we obtain a
corresponding graphical representation of the causal structure over
the vertices $(1, \ldots, d)$ by drawing directed edges from
$\PA[]{k}$ to $k$ for all $k \in \{1, \ldots, d\}$.
We further assume that 
the joint noise distribution is absolutely 
continuous with respect to a product measure
and that it factorizes, i.e., the noise components are assumed to be jointly independent.
As before, 
we require the system~\eqref{eq:scmiid}
to be uniquely solvable, 
which
is always
satisfied if the graph is acyclic, for example. 
An SCM induces a unique joint distribution
over the variables $X_1, \ldots, X_d$ 
\citep[e.g.,][]{Bongers2016b}, and
an observed data set may be modeled as a collection of i.i.d.\ realizations from that
distribution.

The two approaches described above
serve different purposes.
The model described in~\eqref{eq:obsnoiseiid}
might be helpful when 
the underlying system is assumed to be deterministic and all randomness can
be thought of as measurement noise, for example.
While this might be a realistic assumption in many applications,
the approach comes with various 
statistical difficulties,
including the famous errors-in-variables problem
\citep{Carroll2006}
and an increased difficulty when identifying
parameters or causal structure from data \citep{Zhang2018uai}.
We speculate that this is one of the reasons,
why less work seems to be devoted to the first approach.
The second approach is better understood
but assumes 
that 
the noise 
is not purely measurement noise,
but
enters
into the causal
mechanism. 
It depends on the application at hand,
whether this assumption is reasonable.

\subsection{Interventions} \label{sec:interventionsiid}
SCMs allow us to define \emph{interventions}.  
For any
$j \in \{1, \ldots, d\}$, we can replace one of the assignments 
in~\eqref{eq:scmiiddet} or~\eqref{eq:scmiid}.
In the former case, 
we could replace the 
assignment with 
$x^{j} := \tilde{f}^{j}(x^{\tPA[]{j}})$
and in the latter case with 
$X^{j} := \tilde{f}^{j}(X^{\tPA[]{j}}, \tilde{\varepsilon}^{j})$, for example.
Usually, we restrict ourselves to interventions 
that yield 
a new SCM, 
so the interventions must respect unique solvability.
If that is the case, 
the intervention induces a new state of the system, that we denote by 
$do(x^{j} := \tilde{f}^{j}(x^{\tPA[]{j}}))$
or
$do(X^{j} := \tilde{f}^{j}(X^{\tPA[]{j}},
\tilde{\varepsilon}^{j}))$,
respectively. 
An intervention on one of the variables
propagates through the system, possibly  affecting many other variables that are graphical descendants of the targeted node. %
For the stochastic SCMs from Section~\ref{sec:SCM_drivingnoise},
one may think about randomized experiments as a $do$-intervention
and the well-known hard (or point) interventions $do(X^j := 4)$ appear as a special
case. 
\citet{Pearl2009} provides 
many insightful examples of SCMs and interventions throughout his book.  \citet{Bongers2016b}
give measure theoretic details underlying the construction of
SCMs.  
Below, we extend the concept of SCMs to dynamical systems and
give a concrete example 
of an SCM, a graph, and interventions
in that context, see
Figure~\ref{fig:example_ODErep}.

The above definition clarifies which parts of the distribution remain
invariant under interventions.  
In the case of Section~\ref{sec:SCM_drivingnoise},
each conditional distribution $X^k$,
given $X^{\PA[]{k}} = x$, is determined by the structural assignment
for $X^k$.  Thus, two distributions induce the same conditionals
$X^k \,|\, X^{\PA[]{k}} = x$ if one of the distributions is induced by
an SCM and the other one is induced by the same SCM after
intervening on a fixed $j\neq k$, for example.

It may further be instructive to think about equivalence of two causal models. 
They may be called observationally equivalent if they induce the same 
observational distribution and interventionally equivalent 
if they induce the same observational distribution as well as the same intervention distributions \citep[e.g.,][Section~6.8]{Peters2017book}. 
One of the fundamental problems when learning causal structures from data is 
that two causal models may be observationally equivalent, but not interventionally equivalent. 

\subsection{Time dependent data}
In many practical applications, an i.i.d.\ data set 
does not provide an adequate description for 
the data sample at hand.
In particular, the concepts above are lacking the notion of time.
Different causal methodology
and several extensions of SCMs 
have been 
proposed 
\citep{Wiener1956, Granger1969, Schreiber00, White10, Hyttinen2013,
  Peters2013nips, 
  Pfister2018jasa},
mostly 
considering
discrete time models such as vector autoregressive models \citep{Luetkepohl07}, for example.
\citet{Peters2009icml, bauer2016arrow}
discuss the relation between causality and the arrow of time.
Causal inference for longitudinal studies
has been studied extensively, too \citep[e.g.][]{
VanderWeele2015,Aalen2008,RobinsLong},
where the results are often 
formulated in the language of
potential outcomes \citep{Imbens2015}, rather than SCMs.
In this article, we 
focus 
on continuous time systems that are governed by ordinary differential
equations. 
In particular, we propose a natural and straight-forward extension 
of the notion of SCM to dynamical systems.  
The construction closely follows the existing ideas of SCMs 
and interventions. 
Similar constructions have been suggested elsewhere, and we try our best to provide the relevant references and point out existing differences.
Parts of this book chapter are taken from
\citet{Pfister2018arxiv,Pfister2019submitted}, where we focus on model selection and parameter inference.

\section{Chemical reaction networks and ordinary differential equations}

In many natural sciences and even some social sciences, 
there are 
processes that can be modeled 
by a set of  governing differential equations. 
Examples are found in diverse areas such as 
bioprocessing \citep[e.g.][]{ogunnaike1994process}, 
economics \citep[e.g.][]{zhang2005differential},
genetics \citep[e.g.][]{chen1999modeling}, 
neuroscience \citep[e.g.][]{friston2003dynamic} or
robotics \citep[e.g.][]{murray2017mathematical}.
Below, we provide two examples that come from a subclass of dynamical
models, namely those that are driven by chemical reactions and connect
to ordinary differential equation (ODE) based models by mass action
kinetics.  The general principles, however, can readily be extended to
more complex model classes.
Formally, a general reaction \citep[e.g.][]{Wilkinson09} takes the form
\begin{equation*}
  m_1 R_1 + m_2 R_2 + \ldots + m_r R_r \rightarrow
  n_1 P_1 + n_2 P_2 + \ldots + n_p P_p,  
\end{equation*}
where $r$ is the number of reactants and $p$ the number of
products. Both $R_i$ and $P_j$ can be thought of as molecules and are
often called {species}. The coefficients $m_i$ and $n_j$ are positive integers, called
stochiometries. We 
now provide two examples:
(1) a famous and often used model that describes the
abundance of predators and prey, illustrating the law of mass-action
kinetics and (2) Michaelis-Menten kinetics which results in
nonlinear ODEs.
\paragraph{Lotka-Volterra
  model} \label{example:mass_action}
The Lotka-Volterra model \citep{Lotka09} takes the
form 
\begin{align}
  A &\overset{k_1}{\longrightarrow} 2A
      \label{eq:lotkareac1}\\
  A + B &\overset{k_2}{\longrightarrow} 2B
          \label{eq:lotkareac2}\\
  B & \overset{k_3}{\longrightarrow} \emptyset,
\end{align}
where $A$ and $B$ describe abundance of prey and predators, respectively. In this model, the prey reproduce by themselves, 
but the predators require abundance of prey for reproduction.
The coefficients
$k_1, k_2$, and $k_3$ indicate the rates, with which the reactions
happen.
  
\begin{figure}[t]
  \centering
  \includegraphics[width = 0.6\textwidth]{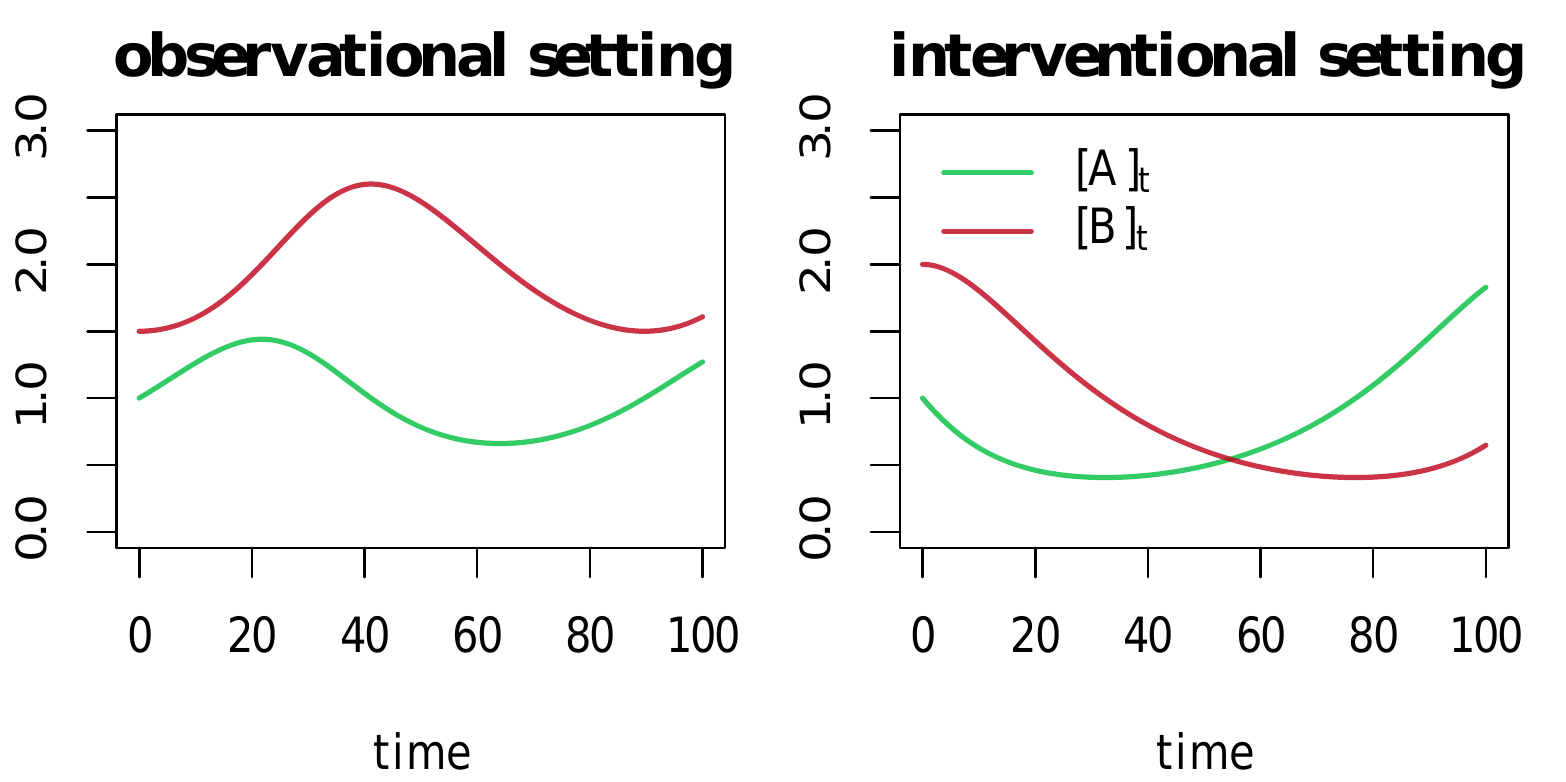}
  \caption{Example trajectories for the basic Lotka-Volterra model
    given in \eqref{eq:lotka1} and \eqref{eq:lotka2}. Left plot
    correspond to observation setting with rates $k_1=0.1$, $k_2=0.05$
    and $k_3=0.05$ and initial values $[A]_0=1$ and $[B]_0=1.5$ and
    right plot to an intervention where we set $k_1=0.05$ and
    $[B]_0=2$.  }\label{fig:examplesinterv}
\end{figure}

In mass-action kinetics \citep{Waage64}, one usually considers the
concentration $[X]$ of a species $X$, the square parentheses indicating
that one refers to the concentration 
rather than to the 
integer number
of abundant species or molecules. The law of mass-action
states that the instantaneous rate of each reaction is proportional to
the product of each of its reactants raised to the power of its
stochiometry. For the Lotka-Volterra model this yields
\begin{align}
  \tfrac{\text{d}}{\text{d}t}[A] &= k_1 [A] - k_2 [A][B] \label{eq:lotka1}\\
  \tfrac{\text{d}}{\text{d}t}[B] &= k_2 [A][B] - k_3 [B] \label{eq:lotka2}. 
\end{align}
Figure~\ref{fig:examplesinterv} shows solutions for these differential
equations for both an observational setting (left plot) with rates $k_1=0.1$,
$k_2=0.05$ and $k_3=0.05$ and initial values $[A]_0=1$ and $[B]_0=1.5$ as well
as an interventional setting (right plot) where we set $k_1=0.05$ and $[B]_0=2$.
Even though equations~\eqref{eq:lotka1} and~\eqref{eq:lotka2}
contain interaction terms
of the concentration of the different species,
they are linear in the model parameters,
a property that is exploited by
many practical methods. 

\paragraph{Michaelis-Menten kinetics}
In Michaelis-Menten kinetics \citep{MM13}, the starting point is a
specific enzyme reaction given by the equations
 \begin{align*}
   E + S &\overset{k_1}{\longrightarrow} ES\\
   ES &\overset{k_2}{\longrightarrow} E + S\\
   ES & \overset{k_3}{\longrightarrow} E + P,
 \end{align*}
 where the enzyme $E$ binds to a substrate $S$ and finally releases a
 product $P$. Under some simplifying assumptions regarding the
 relation of rates of the reactions, this yields the
 equation
 \begin{equation}
   \label{eq:MM}
   \tfrac{\text{d}}{\text{d}t}[P] = c_1 \frac{[S]}{c_2 + [S]},
 \end{equation}
 where $c_1,c_2$ are constants. There are many reactions that can be
 described by this model; \citet{MM13} used it to describe how the
 enzyme invertase catalyzes the hydrolysis of sucrose into glucose and
 fructose.

\section{Causal kinetic models} \label{sec:SDMs}
We now define a causal model class for dynamical systems.
The reader may think about the example of a 
Lotka-Volterra model 
as described in~\eqref{eq:lotka1} and~\eqref{eq:lotka2} or 
Michaelis-Menten kinetics~\eqref{eq:MM},
both of which fit into the general framework described below.
In analogy to
Sections~\ref{sec:SCM_measurenoise} and~\ref{sec:SCM_drivingnoise},
we first consider a deterministic version with measurement noise and
secondly a version where the randomness enters inside the structural
equations.

\subsection{Causal kinetic models with measurement noise}
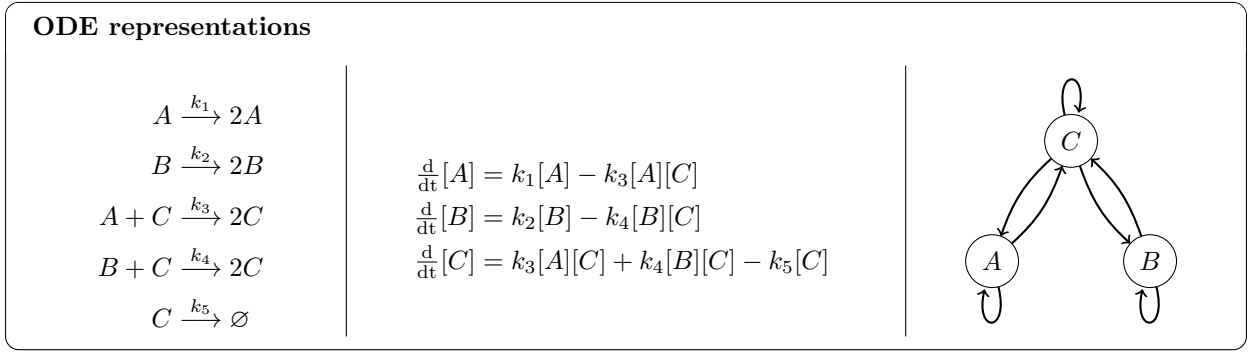
\begin{figure}
  \centering
  \begin{mdframed}[roundcorner=5pt, frametitle={ODE representations}]
    \begin{minipage}[c]{0.25\textwidth}
\begin{center}
      \begin{align*}
        A &\overset{k_1}{\longrightarrow} 2A\\
        B &\overset{k_2}{\longrightarrow} 2B\\
        A + C &\overset{k_3}{\longrightarrow} 2C\\
        B + C &\overset{k_4}{\longrightarrow} 2C\\
        C &\overset{k_5}{\longrightarrow} \varnothing
      \end{align*}
\end{center}
    \end{minipage}
    \hfill\vline\hfill
    \begin{minipage}[c]{0.45\textwidth}
      \begin{align*}
        \tfrac{\text{d}}{\text{dt}}[A] &= k_1[A] - k_3[A][C]\\
        \tfrac{\text{d}}{\text{dt}}[B] &= k_2[B] - k_4[B][C]\\
        \tfrac{\text{d}}{\text{dt}}[C] &= k_3[A][C] + k_4[B][C] - k_5[C]
      \end{align*}
    \end{minipage}
    \hfill\vline\hfill $ $
    \begin{minipage}[c]{0.25\textwidth}
\begin{center}
      \begin{tikzpicture}[xscale=0.7,yscale=0.8]
        \tikzstyle{VertexStyle} = [shape = circle, minimum width = 2em,draw]
        \SetGraphUnit{2}
        \Vertex[Math,L=A,x=-4.5,y=0]{A}
        \Vertex[Math,L=B,x=-1.5,y=0]{B}
        \Vertex[Math,L=C,x=-3,y=2]{C}
        \tikzset{EdgeStyle/.append style = {->, bend right=15}}
        \Edge(A)(C)
        \Edge(C)(A)
        \Edge(B)(C)
        \Edge(C)(B)
        \draw (C) edge [loop above, thick](C);
        \draw (A) edge [loop below, thick](A);
        \draw (B) edge [loop below, thick](B);
      \end{tikzpicture}
\end{center}
    \end{minipage}
  \end{mdframed}
  \caption{Illustration of different ODE representations: 
    (chemical) reactions (left), ODE system derived by mass-action kinetics (middle)
    and corresponding graph (right).}
  \label{fig:example_ODErep}
\end{figure}

We regard the following definition as a natural and straight-forward extension of
SCMs, even though we have not seen 
it in this form 
before.
A \emph{deterministic causal kinetic model} over processes
$\B{x} := (\B{x}_t)_t := (x^1_t, \ldots, x^d_t)_t$ is a collection of
$d$ ODEs and initial value assignments
\begin{align*}
\dott{x}_t^1 &:= f^1(x_t^{\PA[]{1}}), \qquad x^1_0 := \xi^1_0,\\
\dott{x}_t^2 &:= f^2(x_t^{\PA[]{2}}), \qquad x^2_0 := \xi^2_0,\\
& \; \; \; \vdots \\
\dott{x}_t^d &:= f^d(x_t^{\PA[]{d}}), \qquad x^d_0 := \xi^d_0.
\end{align*}
Here, for any $k \in \{1, \ldots, d\}$, $\dott{x}_t^k$ denotes the
time derivative of the component $x^k$ at time~$t$ and
$\PA[]{k} \subseteq \{1, \ldots, d\}$ 
  is called the set of direct parents
of $x^k$ (and may include $x^k$ itself).  We require that the system of initial value problems is
uniquely solvable. For each causal kinetic model, we can obtain a
corresponding graph over the vertices $(1, \ldots, d)$ by drawing
edges from $\PA[]{k}$ to $k$, for $k \in \{1, \ldots, d\}$ (see
Figure~\ref{fig:example_ODErep}).  If we consider the initial values
as random variables, this induces a distribution over
$\B{x} = (\B{x}_t)_t$.

Similarly as in the case of deterministic SCMs, 
{causal kinetic models} are deterministic models describing an
underlying causal structure. 
The observed data can then be modeled as
noisy observations of the system, i.e.,\footnote{Alternatively, one may add the noise variables only at observed time points.}
\begin{equation} \label{eq:noise}
{\B{X}}_t = \B{x}_t + \boldsymbol{\eps}_t,
\end{equation}
where one may assume for simplicity that each noise component of
$\boldsymbol{\eps}_t$ is i.i.d., for example. This induces a distribution
over ${\B{X}} = ({\B{X}}_t)_t$. 

\subsection{Causal kinetic models with driving noise}
As for SCMs, the randomness might 
also be added 
directly into the
structural assignments. This 
yields a more involved mathematical formulation
though, since the objects of interest are
continuous time processes. We define a \emph{stochastic causal kinetic
  model} 
  over processes
$\B{X} := (\B{X}_t)_t := (X^1_t, \ldots, X^d_t)_t$ as a collection of
$d$ stochastic differential equations (SDEs) and initial value
assignments
\begin{equation}
  \label{eq:stochastic_ckm}
  \text{d}X_t^k := f^k(X_t^{\PA[]{k}})\text{d}t + h^k(X_t^{\PA[]{k}})\text{d}W_t^k, \qquad X^k_0 := \xi^k_0,
\end{equation}
where $\text{d}W_t^k$ are independent white noise processes, i.e.,
$W_t=\int_0^t \text{d}W_s$ is a Brownian motion.\footnote{Readers who
  are not familiar with the formal definition of SDEs may think about
  them as a notational abbreviation for the integrated form, i.e.,
  $X_t^k := \int_0^t f^k(X_s^{\PA[]{k}})\text{d}s +
  \int_0^th^k(X_s^{\PA[]{k}})\text{d}W_s^k$.} Again, we require
that the SDEs in \eqref{eq:stochastic_ckm} are uniquely solvable,
which in the setting of SDEs becomes substantially harder to verify.
The functions $f^k$ are called drift coefficients and the functions
$h^k$ are called diffusion coefficients. Intuitively, it can be
helpful to think about the change $X_{t+\Delta}^k-X_t^k$ as being
normally distributed with expectation
$f^k(X_t^{\PA[]{k}})\cdot\Delta$ and variance
$h^k(X_t^{\PA[]{k}})^2\cdot\Delta$. In the most basic setting, $h^k$
can be assumed to be constant which results in an integrated equations
of the form
\begin{equation*}
  X_t^k := \int_0^t f^k(X_s^{\PA[]{k}})\text{d}s + W_t^k.
\end{equation*}
In general, solving SDEs is a difficult problem and numerical
procedures often have slower rates compared with their deterministic
counterparts \citep{han2017random}.  We believe that despite these
difficulties, SDE-based causal models may potentially prove useful in
several areas of applications.
There is some work that has made 
first attempts to circumvent the
difficulties of models using SDEs by looking at random differential
equations \citep{ bauer2017, Bongers2018, abbati2019}, which still
allow including randomness directly into the causal structure.  As
for SCMs, it depends on the application whether a causal model with
measurement noise or the full stochastic setting is the more
appropriate choice.

\subsection{Interventions} \label{sec:interventions}

An intervention on the system replaces some of the structural
assignments. Interventions can change the dynamics of the process
$x^k$, the initial values or both at the same time. 
This definition allows for 
several ways of manipulating the system, 
which may prove useful when modeling
complex dynamical systems and their perturbations; some of the
possibilities are discussed below. Formally, for a deterministic
causal kinetic model over a process $(\B{x}_t)_t$, an
\emph{intervention} on the process $x^k$ for $k\in\{1,\dots,d\}$,
corresponds to replacing the $k$-th initial condition or the $k$-th
ODE with
\begin{equation*}
  x^k_0 := \xi
  \quad\text{or}\quad
  \dott{x}^k_t := g(x_t^{\PA[]{}}),
\end{equation*}
respectively, where $\PA[]{} \subseteq \{1, \ldots, d\}$ is the set of
new parent components.  In both cases, we still require that the
system of initial value problems is uniquely solvable. The
interventions are denoted by
\begin{equation*}
  {\doo{{x}^k_0 := \xi}}
  \quad \text{ and } \quad 
  {\doo{\dott{x}^k_t := g(x_t^{\PA[]{}})}},    
\end{equation*}
respectively. %
The same definitions apply in the presence of observational
noise~$\eps_t$, see \eqref{eq:noise}, where the noise is added after
the system has been perturbed. For a stochastic causal kinetic model,
we analogously define the interventions
\begin{equation*}
  \doo{X^k_0 := \xi}
  \quad \text{ and } \quad 
  \doo{\text{d}X_t^k  := g(X_t^{\PA[]{}}) + j(X_t^{\PA[]{}})\text{d}W_t^k}.    
\end{equation*}
While we regard both deterministic and stochastic causal kinetic
models as potentially relevant for practical applications, we will, in
the remainder of this chapter, focus on deterministic causal models.

If the ODE system is induced by a set of reactions,
a natural class of interventions is described by replacing one (or
some) of the reactions.  In the Lotka-Voltera model from
Section~\ref{example:mass_action}, changing the rate of the first
reaction~\eqref{eq:lotkareac1}, i.e., changing $k_1$ to $\tilde{k}_1$,
say, yields a change of the assignment~\eqref{eq:lotka1}.  Changing the
rate of the second reaction~\eqref{eq:lotkareac2}, however, yields a
change of both assignments~\eqref{eq:lotka1} and~\eqref{eq:lotka2}.
In general, changing one of the reactions induces a change in
differential equations for all variables that appear in the reaction
(either on the left or on the right). The proposed framework
additionally allows us to set a variable $x^k$ to a constant value $c$
by performing the interventions $ \dooo{x^k_0 := c} $ and
$\dooo{\dott{x}^k_t := 0}$.  To obtain a softer version of this
effect, we may also introduce a forcing term that ``pulls'' the variable
$x^k$ to a certain value $c$.  Alternatively, one can keep the
dependence of $\dott{x}^k$ on $x^{\ell}$ intact, but change the strength
of this dependence,
or even completely change the parent set.

We believe that in a system that is described
well by a system of differential equations, it is most natural to
formulate the interventions as differential equations,
too. 
Nevertheless, 
for a differentiable $\zeta$, 
interventions of the
form 
$x^k_t := \zeta(x^A_t)$ with
$A \subseteq \{1, \ldots, d\} \setminus \{k\}$
\citep[e.g.][]{Hansen2014}, and
$x^k_t := \zeta(t)$
\citep[e.g.][]{Rubenstein2016} 
are included in
the above formalism as well.
The intervention $\dooo{x^k_t := \zeta(x^A_t)}$
can be obtained
by $\dooo{\dott{x}^k_t := \frac{d}{dt}\zeta(x^A_t)}$ and
$\dooo{x^k_0 := \zeta(x^A_0)}$.  Similarly, $\dooo{x^k_t := \zeta(t)}$
is realized by $\dooo{\dott{x}^k_t := \dott{\zeta}(t)}$ and
$\dooo{x^k_0 := \zeta(0)}$.

\subsection{Other causal models for dynamical systems and related work}
We introduced the formal framework of causal
kinetic models that allows us to model dynamical systems
with a set of differential equations and specify what we mean
by intervening in the system. 
Several other useful proposals have been made
that connect differential equations with causality.  
Here, we briefly review some of these suggestions and 
point out a few of the differences.
In general, the attempts are tailored towards different goals.

\citet{MooijJS2013, Blom2018, Bongers2018} and \citet{Rubenstein2016}
consider (deterministic and random) ordinary differential equations.  
Their goal 
is to describe the asymptotic solution of such a system as a causal
model.  
The authors consider interventions that fix the full time trajectory
of a variable to a pre-defined solution, e.g., to a
constant. \citet{MooijJS2013} consider interventions on the ODE system
itself. In that work, the authors are primarily interested in the equilibrium of the ODE
system (assuming that it exists) and its relation to standard
structural causal models (SCMs); they
explicitly do not
distinguish between interventions that yield the same equilibrium.
These approaches may be particularly useful when the focus lies on the stationary solution, rather than the full dynamics.
\citet{Hansen2014}
consider stochastic differential equations, which contain ODEs as
a special case, and introduce interventions, for which at any point in time
the intervened variable can be written as a deterministic
function of other variables.

In practice, the application at hand determines which 
of the models and interventions are most appropriate for describing the real world experiment. 
The structure of causal kinetic models closely follows the spirit of
the SCMs described above. In particular, its modular structure once
more highlights which parts remain invariant under interventions.

\section{Challenges in causal inference for ODE-based
  systems}\label{sec:structure_learning}
Formalizing a causal model for dynamical systems can be taken as a
starting point to conduct causal inference.  Similarly to the i.i.d.\
case, we might be interested in adjustment results, do-calculus, the effect of hidden variables, or causal discovery
\citep[see, e.g.,][]{Pearl2009}.
To the best of our knowledge, 
for dynamical systems,
most of
such questions are still open.
Possible reasons are
the difficulties that arise when working with dynamical systems, some of which we highlight below.
(1) In the deterministic settings, solving a standard algebraic
equation is easier than solving an algebraic equation involving
differentials.  (2) When adding observational noise, the induced
distributions on the left hand side of the structural assignments are
more complicated in causal kinetic systems than in SCMs.
(3) Suppose that in the i.i.d.\ case~\eqref{eq:scmiid}, the noise
variables are additive.  If the parents of each variable (and
therefore the structure of the whole system) is known, the causal
mechanisms, i.e., the functions $f^k$, can be estimated by nonlinear
regression techniques.  In contrast, in the case of
dynamical systems, the fitting process is much more involved, and many
different methods have been suggested.  This includes various
versions of goodness-of-fit of the integrated system, nonlinear least
squares methods or gradient matching \citep{
  Bard1974,Benson79,calderhead2009accelerating, varah1982spline,
  ramsay2007parameter, macdonald2015gradient, Dattner2015,
  data2dynamics, Oates2014, wenk2018fast}.
(4) In the i.i.d.\ setting, 
Markov conditions
connect properties of the graph, such as $d$-separation \citep{Pearl2009}, with properties of the joint distribution, such as conditional independence \citep{Lauritzen1996}.
For dynamical models, however,
it is not apparent that conditional independence is the right notion.
For specific model classes, 
there is interesting work 
exploiting the 
concept of local independence
\citep{Schweder70, Mogensen2018, Mogensen2019, Didelez2000phd, Didelez2008}, with several questions
still being open. 
Finally, (5), in most real world systems, 
not 
all relevant variables are observed, which means that they need to be modeled as hidden
variables.  
While in the i.i.d.\ case there is some understanding of
the effects of hidden variables on observed distributions,
on the identification of causal effects and on causal discovery
\citep[e.g.][]{Spirtes2000, Pearl2009, ShpitserHandbook,
  Richardson2002, Hernan2006, Evans2018, Richardson2017, Hyttinen2012,
  Verma1991, Silva2006}, more work is needed in the case of dynamical
systems.

\section{From invariance to causality and generalizability}
In many
real world systems,
the underlying structure is unknown and
needs to be inferred from data.   
That is, for any $k$, 
we do not know which variables are contained in $\PA[]{k}$.
This setting is often referred to as
{structure learning}
or 
{causal discovery}  \citep{Spirtes2000, Pearl2009}.
To state the problem let us assume that the observed data consist of
$n$ repetitions of discrete time observations of each of the $d$
variables $\B{x}$, or its noisy version $\widetilde{\B{X}}$, on the
time grid $\B{t}=(t_1, \ldots, t_L)$.  By concatenating the time
series for the $d$ variables, one may represent the data by an
$n \times (d \cdot L)$ matrix.
%
%
%
%
%
%
%
%
%
%
Several methods have been suggested to solve this task
\citep[e.g.,][]{Oates2014, mikkelsen2017, data2dynamics}, most of which combine
structure learning, i.e., model selection, with a parameter inference
step.  Some methods \citep{Oates2014} explicitly consider the causal
nature of this problem.  We briefly describe below, in a simplified setting, how it is possible
to exploit the invariances induced by the underlying causal kinetic
model for causal discovery.
Assume there is a target process $y := x^1 $, for which the
parents are unknown and of particular interest.  In short, we assume
that each of the $n$ repetitions has been generated by a model of the
form
\begin{equation} 
  \label{eq:ODEY}
  \dott{y}_t = f^y(\B{x}^{\PA[]{y}}_t),
\end{equation}
for a fixed function $f^y$, possibly with additional measurement noise
$\widetilde{Y}_t = y_t+\varepsilon$.  
This assumption holds, for example, 
if the measurements stem 
from an underlying causal kinetic model
under different interventional settings, 
in none of which the variable $y$ has been intervened on.
In practice, the right-hand side of~\eqref{eq:ODEY} is unknown,
and the goal is thus to identify the causal
predictors among the $\B{x}$, i.e., to infer both the parents
$\PA[]{y}$ of $y$ as well as the function~$f^y$.
In
\citep{Pfister2018arxiv}, we propose a procedure that
specifically 
exploits the invariance of~\eqref{eq:ODEY}
to
tackle the problem of structure learning. 
Each of the repetitions is assumed to be
part of an environment or experimental condition. 
We suppose this assignment is known, e.g.,
repetitions $1, \ldots, 6$ are known to belong to experimental condition one, repetitions $7, \ldots, 19$ to condition two, and all remaining repetitions to condition three.
The method then 
outputs a ranking of models (or variables)
by 
trading off predictability and invariance of such models.
In the i.i.d.\ case, trade-offs in a similar spirit have been suggested by
\citet{Rojas2016},
\citet{Magliacane2018}
and
\citet{Rothenhaeusler2018},
for example.

The model~\eqref{eq:ODEY} is
valid independently of interventions on variables other than $y$ and
can thus be 
used for prediction in a new experimental setup,
even if there are large perturbations on the predictors $\B{x}$.  
As a consequence, the 
method proposed in \citep{Pfister2018arxiv}
outputs models that generalize better to unseen experiments, even 
when considering real data from large metabolic network experiments.
This finding 
adds to a recent debate suggesting to add
invariance as a fitting criteria to data science methodology \citep{ScholkopfJPSZMJ2012, yu2013, Peters2016jrssb, Bareinboim2016pnas,
  Meinshausen2016pnas, yu2019}.  
At its core lies the modularity of the structure of the causal model and its implied relation between causality and invariance.

\section{Conclusions}
We have discussed an extension 
of 
structural causal models to systems that are governed by
differential equations.
As in the i.i.d.\ case, 
the models may be 
equipped with
either measurement noise or driving noise, where the
latter case yields the concept of stochastic 
differential equations.
These two model classes, called causal kinetic models, may serve as a starting point for answering
questions 
commonly asked in the field of causal inference. 
Many of such questions are neither fully understood nor answered,
and more work is needed to gather as much understanding as 
we have for i.i.d.\ data.

The mathematical complexity 
of the models poses a challenge when working with 
kinetic models.
Some aspects of causal inference, however, 
may become easier. 
The concept of faithfulness suggests, for example, that in the i.i.d.\ setting, a child of a random variable is predictive for its parent. This assumption seems less justified in the case of dynamical processes.
Also, considering
local independence
and assuming causal sufficiency, Markov equivalence classes contain only a single directed acyclic graph \citep{Mogensen2019}. 
Both of these points may prove to be useful for causal discovery. 
Furthermore, intervening 
on a set of differential 
equations usually affects 
the whole time trajectory. 
Relatively mild interventions may thus carry a lot of information about the causal
structure. This may be particularly
relevant when the available data are not yet sufficient to identify causal mechanisms and additional data have to be collected. 
There is a close connection between experimentation and causal inference \citep{imai2013experimental, Peters2017book};
the selection of measurement readouts, time points or intervention strategies
could guide experimentation
and has the potential to significantly reduce the number of complicated and expensive experiments. 

While there are several differences to the i.i.d.\ case,
causal kinetic models exhibit the same modularity as 
structural causal models. 
As a consequence, invariance ideas can be exploited in a similar way as it is done in the i.i.d.\ case. 
This includes 
methods that trade off invariance and predictability 
to select models that may generalize better to unseen experiments.

\section*{Acknowledgements}
The authors thank S{\o}ren Wengel Mogensen and Niels Richard Hansen for helpful discussions.
JP was supported by a research grant (18968) from VILLUM FONDEN.

\ifthenelse {\boolean{ARXIVVER}}
{
\bibliographystyle{plainnat}
}{}
\bibliography{bibliography}  %

\end{document}